\documentclass[twocolumn,showpacs,preprintnumbers,amsmath,amssymb]{revtex4}

\usepackage{graphicx}% Include figure files
\usepackage{dcolumn}% Align table columns on decimal point
\usepackage{bm}% bold math
\usepackage{epsfig}

\input epsf

\def\plotthree#1#2#3{\centering \leavevmode
    \epsfxsize=0.66\columnwidth \epsfbox{#1} \hfil
    \epsfxsize=0.66\columnwidth \epsfbox{#2} \hfil
    \epsfxsize=0.66\columnwidth \epsfbox{#3}}

\def\be{\begin{equation}}
\def\ee{\end{equation}}
\def\bea{\begin{eqnarray}}
\def\eea{\end{eqnarray}}

\def\cmm2{{\,\rm cm^{-2}}}
\def\cm2{{\,{\rm cm}^2}}
\def\cmm3{{\,{\rm cm}^{-3}}}
\def\gcmm3{{\,{\rm g\,cm^{-3}}}}

\def\fun#1#2{\lower3.6pt\vbox{\baselineskip0pt\lineskip.9pt
  \ialign{$\mathsurround=0pt#1\hfil##\hfil$\crcr#2\crcr\sim\crcr}}}

\def\p3m{P$^3$M}

\def\la{\mathrel{\mathpalette\fun <}}
\def\ga{\mathrel{\mathpalette\fun >}}
\def\fun#1#2{\lower3.6pt\vbox{\baselineskip0pt\lineskip.9pt
  \ialign{$\mathsurround=0pt#1\hfil##\hfil$\crcr#2\crcr\sim\crcr}}}
\begin{document}

\def\affilmrk#1{$^{#1}$}
\def\affilmk#1#2{$^{#1}$#2;}

\def\ucd{1}
\def\uc{2}

\bibliographystyle{prsty}
%\twocolumn[\hsize\textwidth\columnwidth\hsize\csname @twocolumnfalse\endcsname
\title{On Precision Measurement of the Mean Curvature}
\author{Lloyd\ Knox\affilmrk{\ucd}\footnote{email:lknox@ucdavis.edu}
}
\affiliation{
\parshape 1 -3cm 24cm
\affilmk{\ucd}{Department of Physics, One Shields Avenue
University of California, Davis, California 95616, USA}
}
\date{\today}

\begin{abstract}
Very small mean curvature is a robust prediction of inflation worth 
rigorous checking.  Since current constraints are derived from determinations
of the angular-diameter distance to the CMB last-scattering surface, which
is also affected by dark energy, they are limited by our understanding 
of the dark energy.  Measurements of luminosity or angular-diameter 
distances to redshifts in the matter-dominated era can greatly reduce
this uncertainty.  With a one percent measurement of the distance to $z=3$, 
combined with the CMB data expected from Planck, 
one can achieve $\sigma(\Omega_k h^2) \sim 10^{-3}$.
A non-zero detection at this level would be evidence against inflation
or for unusually large curvature fluctuations on super-Hubble scales.
\end{abstract}
  \maketitle
%\narrowtext
%]

\section{Introduction.}

One of the great triumphs of inflation has been the inference from CMB
observations that $\Omega_{\rm tot} \equiv \rho/\rho_c \simeq 1$
\cite{dodelson00,bennett03}.
Here $\rho$ is the mean total density of the Universe today and $\rho_c$ is
the critical density that is a function of the expansion rate.  As always in 
general relativity, matter properties determine metric properties.  
In particular,  from $\rho = \rho_c$ it follows that the mean 
curvature is zero.  

The triumphant verification of the zero mean curvature prediction was
especially rewarding for inflation theorists given the decades of
strong observational evidence that the density of matter (both
baryonic and dark) is only about 1/3 of the critical density; i.e.,
$\Omega_m \simeq 0.3$.  For example, the ratio of dark matter mass to
baryonic mass (inferred to be about 6 from X-ray observations of the
hot baryons in the dark-matter dominated potential wells of galaxy
clusters) combined with nucleosynthesis determinations of $\Omega_b
\simeq 0.05$ \cite{walker91} lead to $\Omega_m \simeq
0.3$\cite{white93}.

We can understand the difference between $\Omega_{\rm tot} \simeq 1$
and $\Omega_m$ as due to an additional component, called `dark energy',
that is causing the expansion of the Universe to
accelerate\cite{riess98,perlmutter99}.  The dark energy, and our lack
of understanding of it, is actually what currently limits the
precision with which the mean curvature is determined.  

Even if we make the strong assumption that the dark energy is a cosmological
constant, it is not possible to separately determine the cosmological constant
and the mean curvature from CMB data alone \cite{eisenstein98,efstathiou99}.
However, with this assumption, measurements to distances in the
low-redshift (dark energy dominated) era, for example as inferred
from supernovae, can be used to break the CMB parameter degeneracy
and thereby allow simultaneous determination of the cosmological
constant and the mean curvature \cite{white98,lineweaver98}.  

If we assume the dark energy is a cosmological constant then
current constraints from WMAP data alone are $\Omega_{\rm tot} =
1.09^{+0.06}_{-0.13}$\cite{tegmark04}.  The main source of uncertainty here is
due to the uncertain value of the cosmological constant.  
Including the power spectrum
of galaxies from SDSS and luminosity distances to SNe Ia
improves the determination of the cosmological constant,
tightening up the curvature constraint somewhat 
to $\Omega_{\rm tot} = 1.054^{+0.048}_{-0.041}$\cite{tegmark04}.  
The best constraint on the mean curvature (once again assuming the
dark energy is a cosmological constant) comes from the distance
determination to $z \sim 0.35$ made possible by the detection of
the acoustic oscillation feature in the galaxy correlation function
\cite{eisenstein05}; they find $\Omega_{\rm tot} = 1.01\pm 0.009$.  
Again with the assumption of a cosmological constant, supernova
data alone can be used to constrain the mean curvature \cite{caldwell04}.

While the determination of the curvature to $\pm 0.01$ is a remarkable
achievement, the assumption of the dark energy as a cosmological constant
is a very strong one.  Indeed, whether the dark energy is a
cosmological constant or something else is perhaps one of the most
important questions in fundamental physics today.  Given the low level
of our theoretical understanding of the dark energy \cite{weinberg89},
we can not draw robust conclusions if they depend on the assumption
that the dark energy is a cosmological constant. Dropping this assumption
would greatly weaken all of the above constraints on the mean curvature.

%Discuss importance of testing the mean curvature prediction
Very small mean curvature is a highly robust 
prediction of inflation.  During inflation the Universe is in 
a nearly time-translation invariant state.  Perfect time-translation 
invariance would mean inflation lasts forever.  With inflation lasting 
forever, the mean curvature is sent to zero.  The near time-translation
invariance is responsible for the near scale-invariance of the power
spectrum of curvature fluctuations produced during inflation.  That current
data show that the power spectrum is
very close to scale-invariant \cite{seljak04}, is evidence that inflation
indeed lasted a long time and therefore that the mean curvature is
very close to zero.  The absence of order unity fluctuations on large scales,
as evidenced by the small anisotropy of the CMB, is further indication that 
inflation lasted for a long time.   

Exactly how small do we expect this mean curvature to be?  Roughly
speaking, we expect the ensemble average of the curvature to be such
that $\Omega_k h^2 \la 10^{-60}$.  However, no observations are
sensitive to this ensemble average.  The best we can do is determine
the mean curvature as averaged over our Hubble volume.  Due to the
nearly scale-invariant spectrum of fluctuations, we expect the
curvature averaged over our Hubble volume to be such that $|\Omega_k
h^2| \sim 10^{-5}$.

Detecting $|\Omega_k h^2| \ga 10^{-5}$ would have important consequences
for our understanding of the early Universe and the origin of all structure.
Within the context of inflation, it would imply unusually large fluctuation
power on super-Hubble scales.  Other probes of super-Hubble scales
\cite{grishchuk78,turner91} are sensitive to 
gradients across our Hubble patch and thus suppressed by 
factors of $(k/H_0)$.  The probe we consider here is sensitive to the average
departure in our Hubble volume of the curvature from its mean value; it is not
suppressed by factors of $k/H_0$.

Although one could design an inflaton field effective potential to
produce extra super-Hubble fluctuation power, it is far from what is
generically expected.  Generically, the fluctuations are
better-described by a power law at earlier times, with departure from
a power law occurring as one approaches the end of inflation.  Within
a broader context, such a detection might be evidence for some
alternative to inflation.

Recent CMB observations have revealed some puzzling properties of the
largest scales
\cite{peiris03,efstathiou04b,schwarz04,deoliveira04,eriksen04b,hansen04,land05,jaffe05}.
These peculiar features may have their origin in systematic error.
Exploring the possibility of a cosmological origin, by acquiring more
and relevant data, is difficult due to the small number of large scale
modes in our Hubble volume.  It is therefore highly desirable to probe
beyond our Hubble volume.  Measuring mean curvature provides us with
such a probe.  By exploiting the prediction of the ensemble average,
the average measured over our local Hubble volume is then a measure of
fluctuations on scales larger than the Hubble radius.

Anthropic arguments can alter ones intuition about the likelihood of
detectably non-zero mean curvature.  If the final epoch of inflation,
prior to the hot big bang, begins from a tunneling event one no longer
needs long inflation to explain homogeneity on large scales.  The
tunneling event itself creates a highly homogeneous open Universe
\cite{coleman80,gott82}.  Inflation following tunneling is a natural
consequence of the string theory landscape which has many metastable
vacua \cite{bousso00,susskind03}.  Adopting a particular prior on the
distribution of inflaton effective potential shapes, and including
anthropically-motivated constraints on the amount of structure growth,
\cite{freivogel05} find a limit on the allowable magnitude of the
curvature comparable to the current observational limit which they
take to be $\Omega_{\rm total} > 0.98$.  Further, they find a
significant probability that the curvature is near the upper-bound,
with 10\% of the probability lying between $1-\Omega_{\rm tot} = 0.02$
and $4\times 10^{-4}$.  Universes with small but non-zero mean
curvature have also been discussed recently in \cite{adler05}.

On the other hand, detection of a {\em positive} mean curvature would
severely challenge this picture of inflation in the string theory
landscape.  

For the above reasons, precision measurement of the mean
curvature is very well motivated.  We therefore consider here the
challenge of increasing the precision.  Given our lack of understanding of the
dark energy, the answer is straightforward: constrain the contribution
from the low-redshift, dark-energy-polluted Universe by directly
measuring it; i.e., measure distances from here into the
matter-dominated era.  In the following we expand upon this idea, work
out the resulting uncertainties in the curvature for given CMB data
and measurements into the matter-dominated era, and discuss how these
distances might be measured.

\section{The Problem}  

Defining $r_s^*$ as the comoving extent of the sound horizon at the time
of last scattering, we can write the angle it subtends as
\be
\theta_s = r_s^*/D_A(z_*) 
\ee
by definition of the angular diameter
distance to redshift $z$, $D_A(z)$\footnote{We implicitly use {\em comoving} angular diameter distances throughout.}
and use of the small angle approximation.
This angular size can be
determined to very high accuracy from analysis of cosmic microwave
background data.  It sets the scale for the acoustic peaks, 
$l_A = \pi/\theta_s$ \cite{hu01a}.  Thus if we can calculate $r_s^*$, and how
$D_A$ depends on curvature, we can determine the curvature.

To see how $D_A$ depends on curvature we turn to
the line-element for the FRW metric:
\be
ds^2 = dt^2 - a^2(t)\left(\frac{dr^2}{1-kr^2} + 
r^2(d\theta^2 + \sin^2\theta d\phi^2)\right)
\ee
where $k=\Omega_k H_0^2$ and $\Omega_k \equiv 1-\Omega_{\rm tot}$.   
The comoving length from the origin to a point with coordinate 
value $r$ is 
\bea
&  =1/\sqrt{|k|}\sinh^{-1}(\sqrt{|k|}r) & (k < 0) \\
l = \int \frac{dr}{\sqrt{1-kr^2}} &  =r &         (k = 0) \\
& =1/\sqrt{|k|}\sin^{-1}(\sqrt{|k|}r) &    (k > 0)
\eea
We can now calculate the angular-diameter distance as a function
of $l$ by recognizing that an object at distance $r$ subtending
an angle $d\theta$ has length $r d\theta$.  Therefore
\be
\label{eqn:dvsl}
D_A = r = l+ kl^3/6
\ee
 to lowest order in $k$.

If we knew the comoving distance to $z$, $l(z)$, and measured
$D_A$ we could solve for the geometry, $k$.  Of course, we do not
know $l(z)$.  We could calculate it though {\em if we knew the matter
content}.  Using the Friedmann equation we find that a photon
suffers a redshift $z$ in the course of traveling a comoving distance
\be
l(z) = \sqrt{\frac{3}{8\pi G}} \int_0^z \frac{dz'}{\sqrt{\rho_{m,0} (1+z')^3 +\rho_{k,0} (1+z')^2 + \rho_x(z')}}
\ee
where we have defined $\rho_{k,0} \equiv -3k/(8\pi G)$.
Thus we see the sensitivity of $D_A(z)$ to the matter content
in addition to geometry.  Allowing arbitrary freedom in $\rho_x(z)$
destroys our ability to use $D_A(z)$ to determine geometry.

\section{A Solution}

Despite
contamination from dark energy, we can use the angular-diameter distance to
the CMB last-scattering surface, which we will now call $D_{OL}$, to determine 
the curvature, 
as long as we can measure some angular-diameter distance, $D_{OM}$, 
to some redshift, in the matter-dominated era, $z_M$.  The measurement
of $D_{OM}$ effectively controls the dark energy contribution 
to $D_{OL}$ so that their difference only depends on the curvature 
and the matter density.  Starting from $l_{OL} = l_{OM} + l_{ML}$
and solving for $k$ we find
\be
\label{eqn:k}
k = 6\left[\frac{D_{OL}-\left(D_{OM}+l_{ML}\right)}{D_{OL}^3-D_{OM}^3}\right]
\ee
to lowest order in $k$ where
\bea
l_{ML} & = & \int_{z_M}^{z_*} dz/H(z)  \\
  & \simeq & \sqrt{\frac{3}{8\pi G\rho_{m,0}}}\left[\left(1+z_M\right)^{-1/2}-\left(1+z_*\right)^{-1/2}\right]
\eea
and in the final line we have assumed only the matter density contributes
to $l_{ML}$.

Our goal now is to understand how well $k$ can be determined assuming
that we have some measurement of $D_{OM}$ with error $\sigma(D_{OM})$
and $D_{OL}$ and $\omega_m$ constrained by CMB measurements.  The
result is displayed in the left panel of Figure 1.  We first discuss the inference from
CMB data and then speculate about how $D_{OM}$ might be measured.  Depending on
how $D_{OM}$ is measured one will get different values for $\sigma(\Omega_k h^2)$
as shown in the center and right panels, to be explained below.

Since $\theta_s$ can be determined
with very high accuracy, the fractional error in $D_{OL}$ is simply equal
to the fractional error in the sound horizon.  The sound horizon
depends on the baryon-to-photon ratio, because of how this affects the sound
speed, and the matter and radiation densities because of how these affect 
the expansion rate.  Assuming the standard radiation content the only
degrees of freedom can be taken to be the baryon density today and the 
matter density today.  A fit of the sound horizon 
at last scattering (defined as the epoch at which the optical depth 
reaches unity) is given in \cite{hu04} as
\be
r_s^*/{\rm Mpc} = 144.4 (\omega_m/0.14)^{-0.252}(\omega_b/0.024)^{-0.083}.
\ee
Here we have used the common notation for densities, $\omega_m = \rho_{m,0}/\rho_{\rm scale}$ and $\omega_b = \rho_{b,0}/\rho_{\rm scale}$ where
$\rho_{\rm scale} \equiv 3 [100 {\rm km/sec/Mpc}]^2/(8 \pi G) = 1.8791 \times 10^{-29}$ g/cm$^3$.

Thus to calculate $k$, we need to know $\theta_s$ (which we will assume
we know perfectly), $\omega_m$, $\omega_b$ and a distance into the
matter dominated era, $D_{OM}$.  With $\omega_m$, $\omega_b$ and $\theta_s$
we can calculate $D_{OL}$, and with $\omega_m$ we can calculate $l_{ML}$.
Thus we have what we need to use Eq.~\ref{eqn:k} to get $k = 
\Omega_k H_0^2$.  To express it in more convenient units, we calculate
$\Omega_k h^2$ where $h \equiv H_0/[100 {\rm km/sec/Mpc}]$.

The matter and baryon densities can be determined from the acoustic
peak morphology \cite{hu01a}.  We took the three independent elements
of their error covariance matrix, forecasted for 4 years of WMAP and 1 year
of Planck, from \cite{bond04}.  For $D_{OM}$ we simply assume a measurement
with some variance $\sigma^2(D_{OM})$.  We discuss these measurements
in the next section.

To calculate the error in $\Omega_k h^2$ we create 1000 realizations
of the error in $\omega_m$ and $\omega_b$ from their assumed error
covariance matrix, assuming a normal distribution, and add these
errors to their fiducial value.  We also create 1000 samples of the
error in the distance to $z_M$ assuming the
distance error is a Gaussian with variance $\sigma^2(D_{OM})$ and add
these errors to our fiducial value of $D_{OM}$.  For
each sample of $D_{OM}, \omega_b, \omega_m$ we calculate 
$\Omega_k h^2$ using Eq.~\ref{eqn:k} and the other equations as 
described above.  The statistical error in $\Omega_k h^2$ is then
taken to be the square root of the variance of our derived $\Omega_k h^2$
values.  

\begin{figure*}[htbp]
\label{fig:result}
  \begin{center}
    \plotthree{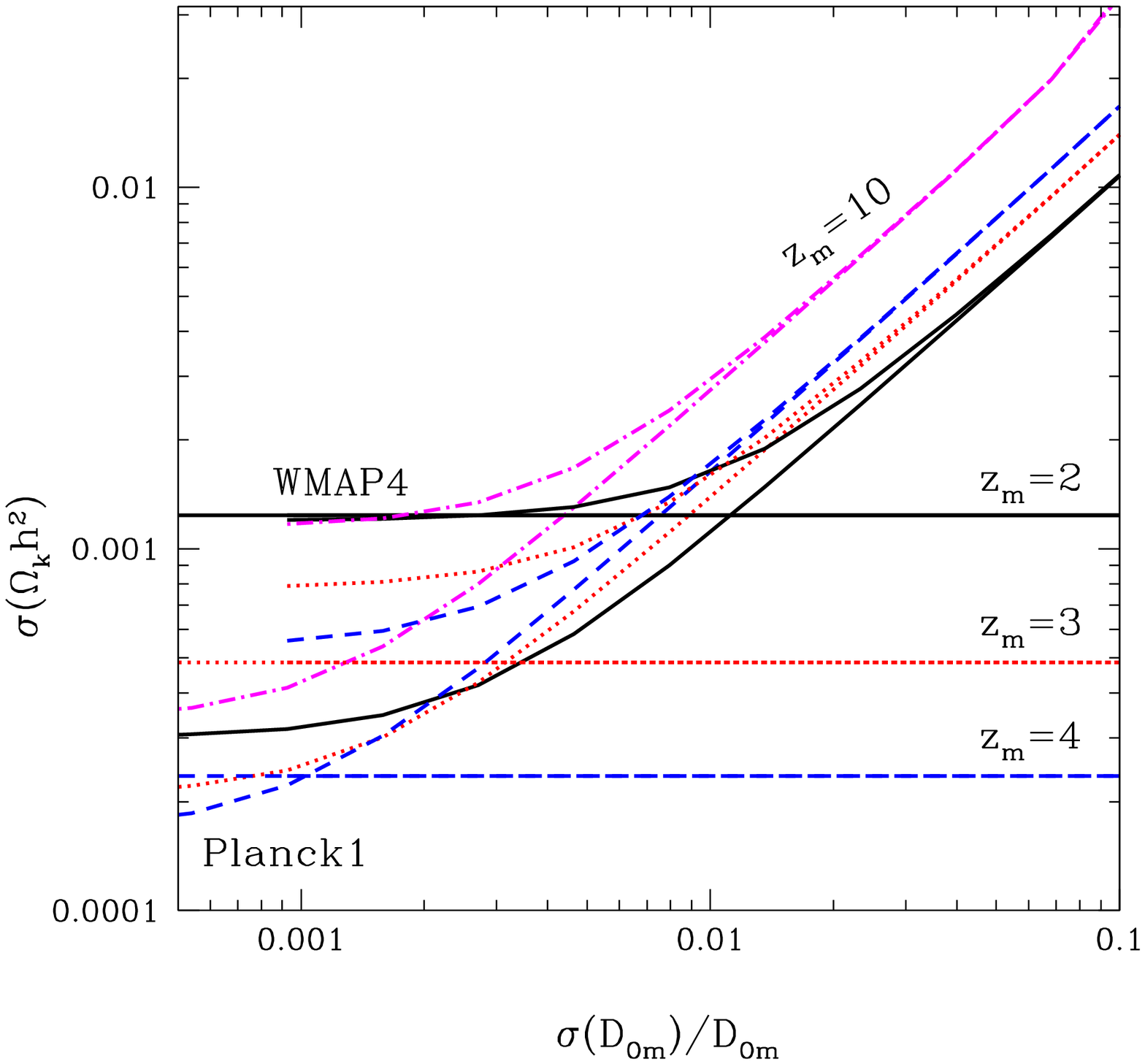}{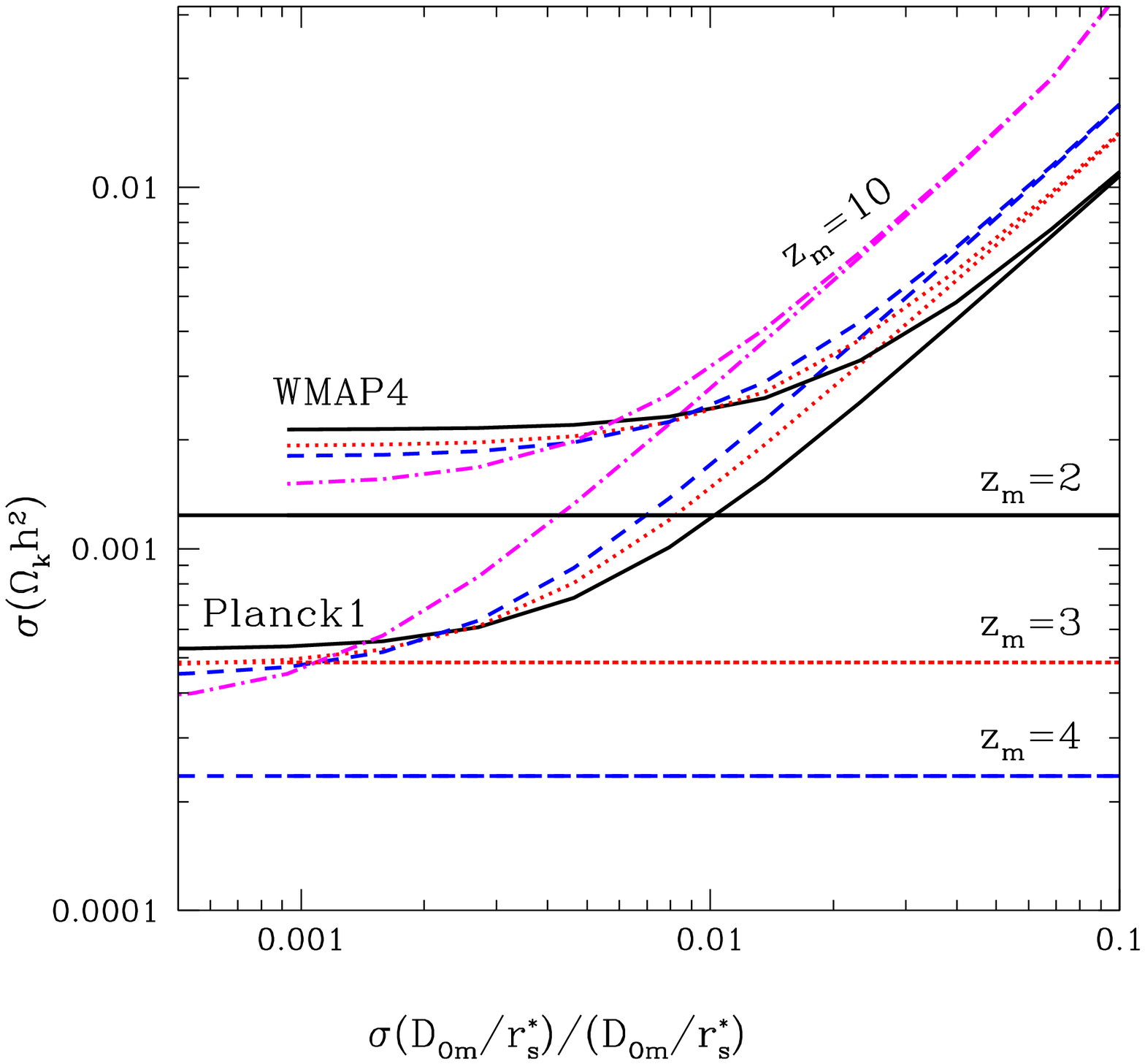}{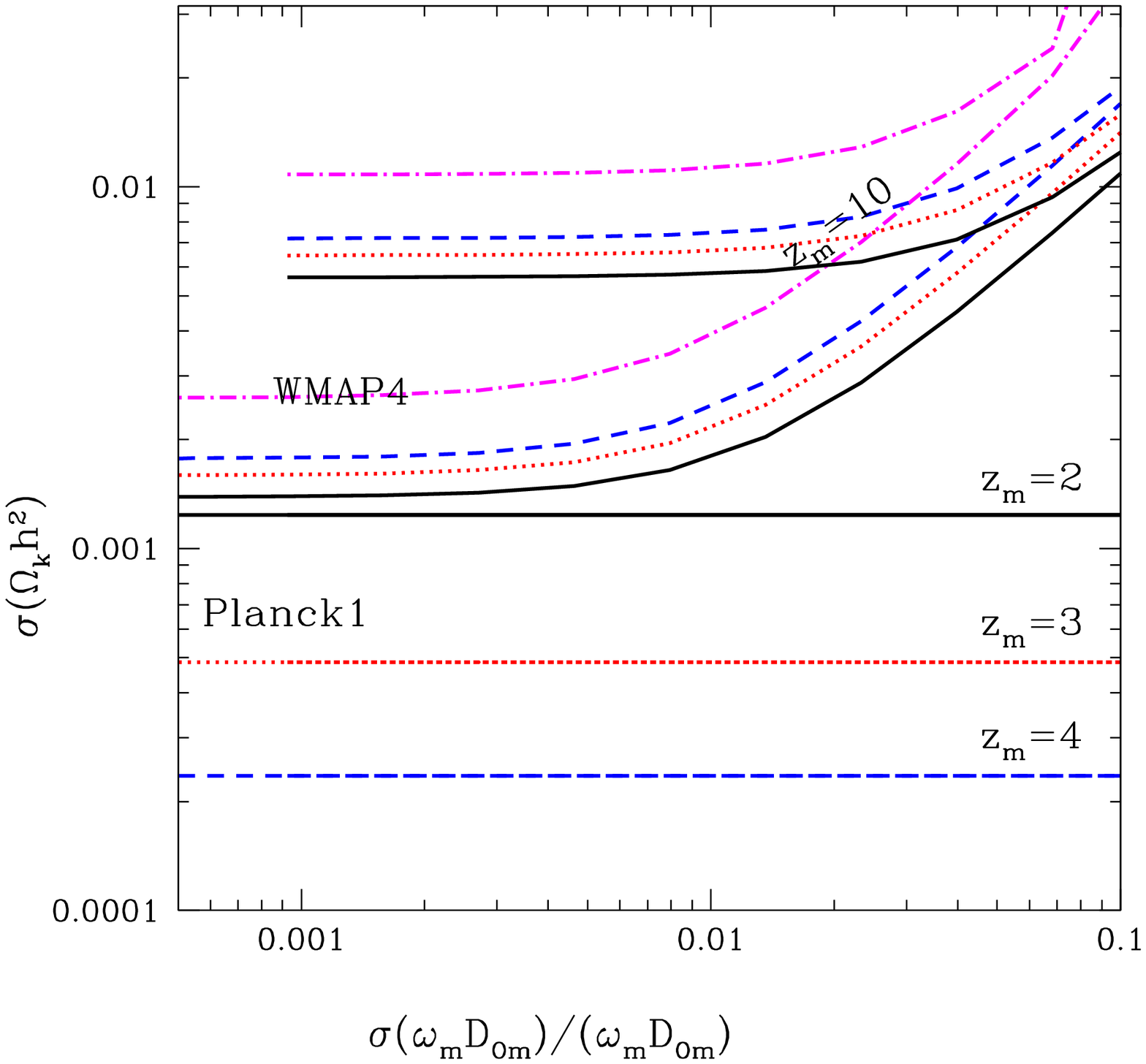}
    \caption{Uncertainty in $\Omega_k h^2$ for {\it WMAP} (higher curves)
and {\it Planck} (lower curves) as a function of the fractional error in
$D_{0M}$ (left panel), $D_{0M}/r_s^*$ (middle panel), and 
$\sqrt{\omega_m}D_{0M}$ (right panel) 
for $z_M = $ 2 (solid), $z_M=3$ (dotted), 
$z_M=4$ (dashed) and $z_M=10$ (dot-dashed).
The straight lines show the amount of error resulting from neglecting
the dark energy contribution at $z > z_M$ assuming our fiducial
model with the dark energy as a cosmological constant.  
}
\end{center}
\end{figure*}

Of course the strategy depends on there being a redshift range $z_M <
z < z_*$ during which the contribution of dark energy to comoving
distances is negligible.  To estimate the level of contamination one
expects from dark energy at $z > z_M$, we have calculated it for the
case of a cosmological constant and plotted the resulting systematic
error in $\Omega_k h^2$ as the horizontal lines in Fig. 1.  Since
our fiducial model was a cosmological constant, the systematic error
is simply $\Omega_k h^2$ averaged over all the samples.

For constant $w$ models consistent with current data, taking $w > -1$
will increase the level of contamination slightly.  More worrisome are
models with more complicated time-dependence, such as the oscillating
model of \cite{dodelson00b}.  Unexpectedly large contamination by dark
energy in the $z_M < z < z_*$ redshift range can be guarded against by
measuring the growth of the matter power spectrum from last-scattering
to $z= z_M$.  If dark energy is making a significant contribution to
the expansion rate in this range, then it will suppress growth.
  
As seen in Fig. 1, at high $\sigma(D_{0M})$ the statistical 
error increases as $z_M$ increases.  We expect this since in 
the limit that $z_M = z_*$, the redshift of 
last-scattering, our measurement of $D_{OM}$ brings us no new
information.  At lower values of $\sigma(D_{0M})$ the trend with
$z_M$ is more complicated due to a cancellation between the $D_{OL}$ error
and the $l_{ML}$ error.  In the limit
that the only uncertain parameter is $\omega_m$ we have
\be
\label{eqn:omegam_limit}
\sigma(\Omega_k h^2) = 6\frac{h^2/H_0^2}{D_{OL}^3-D_{OM}^3} \left(a_1 D_{OL} + a_2 l_{ML}\right) 
\frac{\sigma(\omega_m)}{\omega_m}
\ee
with $a_1 = -0.252$ and $a_1 = 0.5$.  In this limit, the error goes to zero at
$D_{OL} =-a_2/a_1 l_{ML} \simeq 2 l_{ML}$.  

The chief benefit of increasing $z_M$ is reduced systematic
error from dark energy at $z > z_M$.  The statistical error is also
smaller for one year of Planck than for four years of WMAP as long
as $D_{OM}$ is measured well enough that uncertainties in the matter
and baryon densities are important.  

As we will discuss in the next session, a variety of techniques can be
used to get distances from standard rulers in the matter power
spectrum.  These standard rulers are $r_s^*$ and the size of the
comoving horizon at matter-radiation equality, $r_{\rm EQ} \propto
\omega_m^{-1}$.  The measurements thus determine the combinations
$D_{OM}/r_s^*$ and $\omega_m D_{0M}$.  Thus we also plot in the middle
and right panels of Fig. 1 results from
two more calculations, just like the first one, except we assume
independent errors in $D_{OM}/r_s^*$ and then $\omega_m D_{0M}$.
These assumptions lead to correlations between the $D_{OM}$ error and
the $D_{OL}$ error, and therefore we get different results when the
$D_{OL}$ errors are significant.

Although the above analysis only assumes a measurement to one
redshift, any particular methodology for distances measurements will
result in distances to a series of redshifts, some of which may be
heavily contaminated by dark energy.  The actual analysis of the data
will involve simultaneous fitting for dark energy and mean curvature.
The calculations here are done in two illustrative limits:  no dark
energy at $z > z_m$ and completely neglected cosmological constant
at $z > z_m$.  Any conclusions about mean curvature from
future precision measurements will be complicated with arguments about
the possibility of residual amounts of dark energy.  Going to high
redshifts mitigates this confusion greatly, but does not completely
remove it.  Thus distances to more than one redshift in the matter-dominated
era will be useful.

Using our idealized calculation as a rough guide though, it appears
that distance measurements in the $z \sim 3$ to $z \sim 4$ range will be most
helpful.  Lower redshifts are too susceptible to errors arising from
modeling the dark energy (estimated by the horizontal lines in the 
figures) and the statistical errors for $z_M > 4$ become large
since they must diverge as $z_M \rightarrow z_*$, the redshift
of last scattering.. 

Another solution was recently proposed in \cite{bernstein05} that
exploits the fact that $r_{ML}$ is not the angular-diameter
distance from $z_M$ to $z_L$.  Thus, although $r_{OL} = r_{OM} + r_{ML}$,
$D_{OL}- (D_{OM}+ D_{ML}) \propto \Omega_k$.  
Cosmic shear observables are sensitive to all three of these distances.

\section{Distance Measurements}

We have provided motivation for measurement of distances into the
matter-dominated era.  
In this section we briefly discuss different methods for
obtaining these distances.  

\subsection{Supernovae}
The distance-redshift relation at $z \la 1$ is currently determined
best from observations of Type Ia supernovae \cite{riess04}.  Although
these are luminosity distances, in the absence of unknown scattering
or absorption effects, $D_L = (1+z)^2 D_A$.  However, the prospects for
percent level determination of distances at $z \ga 2$ are not good.
Difficulties arises from the $(1+z)^4$ cosmological dimming and the
challenge of determining the amount of reddening (in order to correct
for dust extinction in the host galaxy) with light that was bluer in
the rest frame than is the case for lower $z$ supernovae.  Further,
gravitational lensing of the light from the supernovae will add
additional dispersion to the observed fluxes, increasing the number
needed to obtain a given level of precision
\cite{frieman97,holz98a,holz98b,wang99}.  Finally, current supernova
constraints on $\Omega_m$ and $\Omega_\Lambda$ come from distance
ratios, rather than absolute distance determinations, due to the
(nearly) constant but unknown luminosity of the standard candle.  The supernova
absolute luminosity calibration would have to improve dramatically
in order for per cent level determination of absolute distances.  A sufficient
calibration would be possible with a 1\% determination of $H_0$, which
may be achievable with Square Kilometer Array observations of water 
masers \cite{greenhill04}.

Linder \cite{linder05d} has recently considered the impact of allowing
for non-zero curvature on the ability of a space-based supernova
mission such as SNAP (combined with CMB data) to constrain dark energy
parameters.  He assumes a distribution of supernovae in the interval
$0 < z < 1.7$.  He finds that dropping the assumption of zero mean
curvature greatly weakens the constraints on $w_0$ and $w_a$ when it
is assumed that the equation of state parameter as a function of scale
factor is given by $w(a)=w_0+(1-a)w_a$.

Experiments will not just determine the distance to a single redshift,
as assumed above, but to a range of redshifts.  Thus it is interesting
to see how an actual experiment, making these multiple distance
measurements, can constrain the curvature.  Since the supernovae
measurements do not go beyond $z=1.7$, we expect the results to be
dependent on assumptions made about the dark energy.  Indeed, Linder
\cite{linder05d} finds this to be the case.  If one assumes $w_a = 0$
then $\sigma(\Omega_K) = 0.011$ but if one allows for non-zero $w_a$
(still assuming the form for $w(a)$ above) then the curvature error
weakens by more than a factor of four to $\sigma(\Omega_K) = 0.047$.
We see here in this result a quantification of the degradation of
curvature constraints due to uncertainty about the dark energy, as
expected qualitatively from the discussion and idealized calculations
above.

\subsection{Cosmic Shear}
The statistical properties of cosmic shear are sensitive to both the
distance-redshift relation and the growth of the matter power spectrum
as a function of redshift.  Of course they are also sensitive to the
shape and amplitude of the primordial power spectrum, the matter
density and the baryon density.  With these parameters constrained by
CMB observations, and with sufficient knowledge of the redshift of the
source galaxies (such as from photometrically-determined redshifts)
one can use cosmic shear data to simultaneously reconstruct distance
as a function of redshift and growth as a function of redshift \cite{song05}.  
The combination of Planck's measurement of the CMB and a deep multi-band
ground-based survey of half the sky, such as planned with LSST, 
can determine the distance to
$z=3$ with an error of about 1\% \cite{knox05b}.  The errors in the
$z=3$ measurements are highly correlated with the errors in the $z< 1$
measurements.  The combination of the cosmic shear data with
low-redshift distance measurements (for example, from supernovae) can
therefore improve the the $z=3$ distance determination.

The standard ruler that allows for cosmic shear data to be sensitive
to the distance redshift relation is the turnover in the matter power
spectrum at the comoving size of the horizon at matter-radiation
equality \cite{knox05b} which is proportional to $1/\omega_m$.  The
distance determination is only possible to the extent that $\omega_m$
has been determined.  The errors in the distance will thus be
correlated with errors in $\omega_m$ and therefore with the errors in
$D_{OL}$ and $l_{ML}$.  Errors in the product $D_{OM} \omega_m$ will,
in contrast, be only very weakly correlated with those in $D_{OL}$ and
$l_{ML}$.

The error in $\Omega_k h^2$ in the limit of perfect knowledge of
$\omega_b$ and a perfect measurement of $D_{OM} \omega_m$ (instead of
$D_{OM}$) is again give by Eq.~\ref{eqn:omegam_limit} but now with
$a_1 = 0.748$ and $a_2=-0.5$.  The increased magnitude in $a_1$ means
a larger contribution from the error in $D_{OL}$ and the impossibility
of any significant cancellation with the error in $l_{ML}$.  As a
result, one can see in the right panel of the figure at low values of
$\sigma(D_{OM} \omega_m)/(D_{OM} \omega_m)$ the greatly increased
errors in $\sigma(\Omega_k h^2$) compared to the case of the left
panel.

\subsection{High-$z$ Galaxy Power Spectra}
Acoustic oscillations prior to recombination create a feature in the
matter correlation function with a length scale of $r_s^*$ 
\cite{sakharov65,peebles70,bond84,holtzman89,hu96}.  This feature
can be used as a standard ruler to infer distances from measurement
of $D_A/r_s^*$ \cite{eisenstein98,blake03,linder03,hu03b,seo03,angulo05}
and has recently been used to do so \cite{eisenstein05,cole05}.   
Seo \& Eisenstein \cite{seo03} find
that a photometric redshift survey with redshift errors $\sigma_z$
over a survey spanning $z=2.5$ to $z=3.5$ with solid angle $\Omega$
could achieve an angular-diameter distance determination to $z=3$ of
\be
\frac{\sigma(D_{OM})/r_s^*}{D_{OM}/r_s^*}=.01\sqrt{\frac{2000 {\rm \ sq. \ deg.}}{\Omega}}
\sqrt{\frac{\sigma_z/(1+z)}{0.04}}.
\ee
With spectroscopic redshifts, clustering in the redshift direction
can also be used to determine $H^{-1}(z)/r_s^*$ \cite{seo03} and 
thereby provide a check on the prediction that $H(z) = 8\pi
G \rho_{m,0}(1+z)^3$ where $\rho_{m,0}$ is determined by the CMB
observations.

Achieving a 1\% determination of $D_{0M}/r_s^*$ to $z_M =3$ would
reduce the statistical error on the $\Omega_K$ determination, as reported 
in \cite{eisenstein05}, by a factor of 10.  It would also {\em greatly}
reduce the level of systematic error due to their assumption
of a cosmological constant to a level roughly about that of the
dotted horizontal lines in the figures.  

Using baryon oscillations to determine distances also mitigates a
potential source of systematic error\footnote{I thank G. Bernstein
for suggesting that the baryon oscillation distance measurement's dependence
on CMB calibration might be advantageous.}.  Rather than $\rho_{m,0}$, the
quantity well-determined from CMB observations is the redshift of 
matter-radiation equality, $z_{\rm EQ}$,  because of how it
affects the evolution of gravitational potentials (for a review, see
\cite{hu02e}).  If there were a non-standard radiation content, 
then $\rho_{m,0}$ might be erroneously inferred from $z_{\rm EQ}$.  This
would then lead to an error in $l_{ML}$ and therefore possibly a non-zero
$k$, by Eq.~\ref{eqn:k}, even for a flat cosmology.  

However, as pointed out in \cite{eisenstein04}, CMB observations
robustly determine the combination $\sqrt{\rho_{m,0}} r_s^*$ {\em
independent} of the value of $\rho_{m,0}$.  Therefore, since
baryon oscillations are sensitive to $D_A/r_s^*$, they robustly
determine $\sqrt{\rho_{m,0}}D_A$.  A check for non-zero
$\sqrt{\rho_{m,0}}D_{\rm 0L} - \left(\sqrt{\rho_{m,0}}D_{\rm
0M}+\sqrt{\rho_{m,0}}l_{\rm ML}\right)$ is thus a robust
check for non-zero curvature.  Note that $\sqrt{\rho_{m,0}}l_{\rm ML}$ 
has no cosmological parameter dependence (assuming complete
matter domination); in particular, it does not depend on $\rho_{m,0}$.

\subsection{21cm Radiation from Intergalactic Neutral Hydrogen}
Perhaps the best prospect for measuring distances to redshifts very
deep into the matter-dominated era comes from fluctuations in the
brightness temperature of 21cm radiation from neutral Hydrogen prior
to the complete reionization of the intergalactic medium\footnote{I
thank N. Dalal for suggesting this to me.}\cite{scott90}.  By requiring
statistical isotropy of the fluctuations, in particular that
correlation lengths along the line of sight are equal to correlation
lengths perpendicular to the line of sight, one determines the ratio
$D_A(z)/H^{-1}(z)$ \cite{alcock79,saiyad05}.  With $H(z)$ determined
from the CMB then this can be converted to a measurement of
$D_A(z)$. The $D_A(z)$ might also be determined from observing baryon
oscillations in the 21cm power spectra\cite{barkana05}.  The meter
wavelength signals from redshifted 21cm radiation are much smaller
than contamination from a number of other astrophysical sources, e.g.
\cite{dimatteo02,oh03}.
Quantitative studies show a good prognosis for the ability to clean
out these foregrounds based on their high coherence across frequency,
e.g. \cite{zaldarriaga04,pen04,santos04}.

Such high-z measurements would have the benefit of very low dark 
energy model-dependence; e.g., the horizontal $z_M = 10$ curve is off-scale 
low.  Unfortunately, these measurements are restricted to redshifts
in the pre-reionization era ($z_M> 6$) which have the drawback of larger
statistical errors as discussed above.

\section{Discussion} 
Given the prediction of zero mean curvature, we can use any detection
of curvature averaged over our Hubble volume as evidence of curvature
fluctuations on even larger length scales.  Indeed, inflationary
models predict the existence of these fluctuations, since inflationary
models predict a spectrum of nearly scale-invariant fluctuations.
These predictions are consistent with determinations of the
sub-horizon scale power spectrum from which we infer an rms amplitude
of about $10^{-5}$.  

We have not calculated the dependence of the mean curvature in
our Hubble volume, as measured by the means described above, on
the power spectrum on super-Hubble scales.  This would be interesting
to do, to more completely understand the implications of a non-zero
determination of mean curvature.  A useful starting point for such
a calculation can be found in \cite{sasaki87} and most recently in
\cite{barausse05} who consider spatial fluctuations in the
luminosity distance due to scalar perturbations. 

\section{Conclusions} 
We have emphasized the importance of testing the robust prediction of
inflation that the mean curvature is zero.  We have pointed out that
it is our uncertainty in the dark energy that limits our current 
determinations of the mean curvature and that precise measurements
of the distance to redshifts in the matter-dominated era can circumvent
this problem.  Thus, measurements of the distance-redshift relation
are not only probes of the dark energy, but also of inflation.  

Important as some experiments are that expect null results, it is
always attractive to have a non-zero signal to chase.  It may actually
be possible to reach the level of precision necessary to see a signal
from super-Hubble curvature fluctuations.  Unfortunately this prospect
is a long shot since the signal will have to be two orders of
magnitude larger than naive expectations.  Such a detection would be a
unique datum on these largest scales, allowing us to reach back a bit
further toward the onset of inflation.

\begin{acknowledgments}
I thank A. Albrecht, N. Dalal, N. Kaloper, E. Kolb, S. Perlmutter, 
M. White and H. Zhan for useful conversations.  This work was supported by
the NSF under Grant No. 0307961 and NASA under grant No. NAG5-11098

\end{acknowledgments}

\bibliography{/work3/knox/bib/cmb3}

\end{document}